\documentclass{elsarticle}
\usepackage{amsmath}
\usepackage{graphicx}
\newcommand{\be}{\begin{equation}} 
\newcommand{\ee}{\end{equation}} 
\newcommand{\bea}{\begin{eqnarray}} 
\newcommand{\eea}{\end{eqnarray}} 

\newcommand{\bew}{\begin{widetext}} 
\newcommand{\eew}{\end{widetext}} 
\newcommand{\pd}{\partial} 
\newcommand{\vm}{\vec{m}}
\newcommand{\vmp}{\vec{m}^{'}} 
\newcommand{\vp}{\vec{P}} 
\newcommand{\vx}{\vec{x}}                                                                                                                                                                                                                                                                                                                                                                             
\newcommand{\vq}{\vec{q}}
\newcommand{\epsi}{\epsilon} 
\newcommand{\mc}{\mathcal}
\newcommand{\figcant}
{\begin{figure}[htbp]
        \centering
        \includegraphics[angle=0,width=7cm]{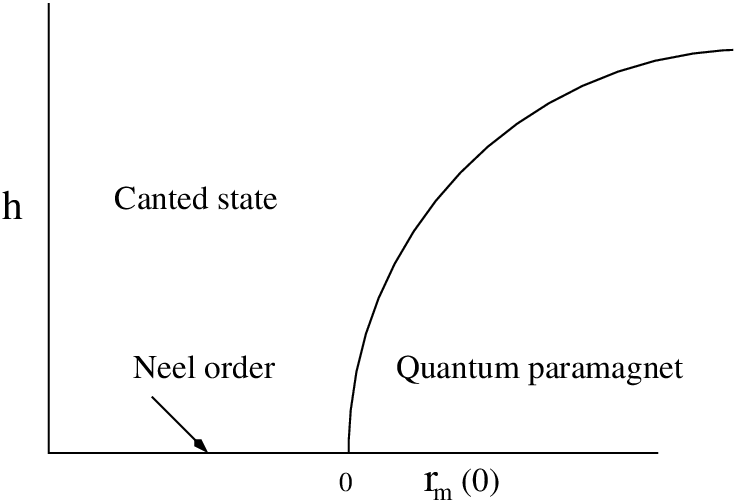}
		\caption{Figure shows the schematic phase diagram for a field induced transition in insulating Heisenberg anti-ferromagnet at zero temperature. At $h=0$, $r\le0$ indicates a Neel order and $h\neq                                                                                                                                                                                                                                                                                                                                                                        0,\, r<0$ region represents Canted state with both ferromagnetic and anti-ferromagnetic order\cite{Sachdev}.}
\label{cant}
	\end{figure}
}
\newcommand{\figchi}
{\begin{figure}[htbp]
        \centering
        \includegraphics[angle=0,width=7cm]{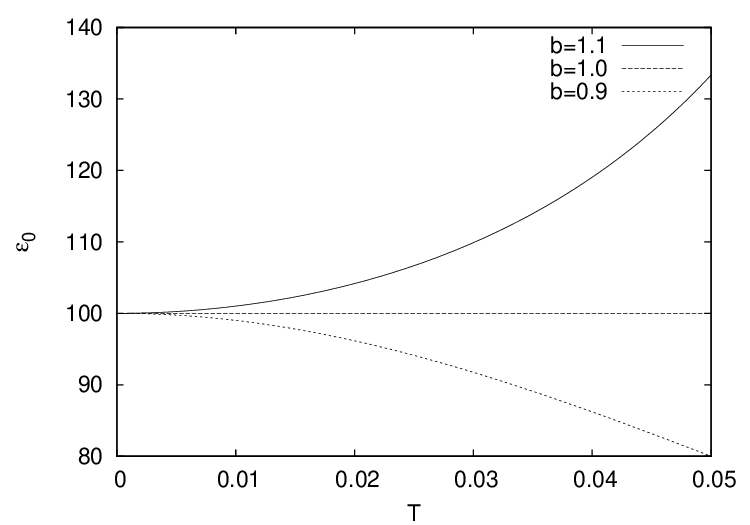}
		\caption{Figure shows the temperature dependence of static dielectric susceptibility near both the anti-ferromagnetic  and ferroelectric quantum critical point for various values of the parameter $b=\frac{\gamma}{\beta}$.
                   Here $\alpha=0.001$ and $\epsilon_0$ and $T$ are plotted in arbitrary scale.}
\label{chi}
	\end{figure}
}
\newcommand{\figchic}
{\begin{figure}[htbp]
        \centering
        \includegraphics[angle=0,width=7cm]{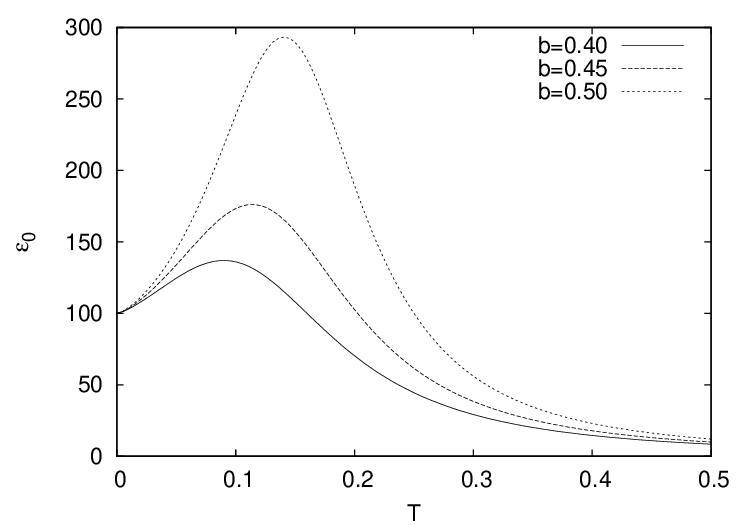}
		\caption{Figure shows the temperature dependence of static dielectric susceptibility near both the anti-ferromagnetic and ferroelectric quantum critical point for various values of the parameter $b=\frac{\gamma}{\beta}$.
                   Here $\nu=3/2,\,{\rm and}\,\alpha=0.01$ and $\epsilon_0$ and $T$ are plotted in arbitrary scale.}
\label{chic}
	\end{figure}
}
\begin{document} 
\title {Quantum critical behavior of a magnetic quantum paraelectric}
\author{Nabyendu Das}

\address[rvt]{Institute of Physics, Bhubaneswar 751005, India.}
\ead{nabyendu@iopb.res.in}

\begin{keyword}
Quantum phase transition \sep Multiferroics \sep EuTiO3
\end{keyword}
\begin{abstract}
We discuss the interplay between anti-ferromagnetic order and polarization fluctuations in a {\it magnetic quantum
paraelectric}. Using an action where anti-ferromagnetic order parameter couples to the polarization fluctuations and 
as well as magnetic field, we derive a set of self consistent equations to study both the temperature and the magnetic 
field dependence of the static dielectric susceptibility. The temperature dependence of dielectric susceptibility near 
both the anti-ferromagnetic quantum critical point and ferroelectric quantum critical point are described using scaling 
arguments. Discussions on achieving various quantum critical points in experiments are also made.
\end{abstract}
\maketitle   
\section{Introduction}
EuTiO$_3$ can be considered as a good addition to the list of perovskite materials exhibiting macroscopic quantum phenomena in ferroelectrics and multiferroics. As far as the structure and the gross features of static dielectric behavior are concerned, this material is similar to the prototype quantum paraelectrics with perovskite structure, like SrTiO$_3$, KTaO$_3$, etc. All these materials are incipient ferroelectric, i.e. they do not undergo any ferroelectric transition at finite temperature, rather their static dielectric susceptibilities reach a very high, temperature independent value ($\epsi_0\sim \mc{O}(10^{4})$ for SrTiO$_3$) at low temperature. Ferroelectric transitions in materials which are similar in structure, for example in BaTiO$_3$, is of displacive type and is well described by softening of a zone center transverse optical mode at some finite temperature. This scenario, that is, the absence of transition and the saturation at high dielectric susceptibility, has been attributed to the 
smallness of the zero temperature gap in the corresponding optical branch. In recent years there is a revival of
theoretical discussions on these materials in context of quantum phase transitions. Recently these materials
have been studied in a perspective of quantum phase transition and fluctuations around it\cite{I, Palova, Rowley, Millis, II}. In the study of quantum criticality in ferroelectrics, EuTiO$_3$ certainly adds a new dimension. This material contains magnetic Eu with spin 7/2, which undergoes anti-ferromagnetic order at $T_N\sim 5.3K$\cite{g-type}. Magnetic order in this material couples to polarization fluctuations and produces a sharp decrease in the static dielectric susceptibility below the Neel temperature. Electric polarization in this material is due to the variations of Ti-O bond-lengths from their equilibrium values. However the collective behavior of such interacting stretched bonds does not lead to a ferroelectric state even at zero temperature. At the critical value of the external magnetic field $\sim 1 {\rm Tesla}$, which suppresses the effects of the Neel order completely, the static dielectric constant of this material attains a quantum paraelectric behavior,  with  $\epsi_0\sim \mc{O}(10^{2})$ 
at zero temperature\cite{Takagi}. The dielectric susceptibility starts getting saturated at a crossover temperature (as defined in ref. \cite{I}) $\sim 30K$. In the present work, we consider the ferroelectric sub system as displacive type, i.e. corresponding order parameter ($\vp$) fluctuations are  represented by the fluctuations in two transverse optic branches. Ferroelectric transition in this case is due to the softening of the optic phonons at the zone center. The effects of dipolar interaction is considered as the stiffening of longitudinal branch, and thus longitudinal fluctuations are not taken into account. Magnetic sector, i.e. the collection of interacting Eu spins $(\vec{S})$ in a cubic perovskite environment and in absence of any external magnetic field, is dominated by anti ferromagnetic fluctuations. Corresponding order parameter ($\vm$) is a vector spin with three components with short range interactions. The effects of the crystal anisotropy are not attempted here, thus all the quartic terms 
are taken as isotropic. We consider a coupling of the form $- \frac{w}{2}|\vp|^2|\vec{S}|^2$ with coupling constant $w>0$, between them and focus on the dependencies of the static dielectric susceptibility on temperature and external magnetic field below $T_N$. In the previous theoretical works\cite{Zhong, wu}, the effects of magneto-electric coupling on the thermodynamic behavior of this material are described at mean field level. But no attention is paid on the possible quantum critical behavior of this system. We will try to explore the behavior of the static dielectric constant in the theoretically interesting anti-ferromagnetic quantum critical point and ferroelectric quantum critical point regime, and discuss the possibility of realizing such a limit in experiments, in the following sections.
\section{Theory}
In the vicinity of quantum critical points, a Landau-Ginzburg-Wilson action for a magnetic quantum paraelectric system, in terms of the sub-lattice magnetization $\vm$ and the electric polarization $\vp$ (soft mode coordinates) can be written in the following form, 
\begin{eqnarray}
\mathcal{A}&=&\int d^dx \int^{\beta}_0 d\tau [(\pd_\tau \vp)^2+(\pd_\tau \vm -i\vec{h}\times \vm)^2\nonumber\\
&+& \frac{c_e}{2} (\nabla\cdot\vp)^2 +\frac{c_m}{2} (\nabla\cdot\vm)^2 + \frac{r_e}{2} (\vp\cdot\vp)\nonumber\\
&+& \frac{r_m}{2} (\vm\cdot\vm) +\frac{u}{2}(\vp^2)^2+\frac{v}{2}(\vm^2)^2\nonumber\\
&-&\frac{w}{2}|\vp|^2|\vm|^2]
\label{S}
\end{eqnarray}
Here $\tau$ is the imaginary time, $\beta$ is the inverse temperature, and $\vec{h}$ is the applied external uniform magnetic field and  $r_e$ and $r_m$ are the non-thermal parameters which
can be tuned to zero to have ferroelectric and anti-ferromagnetic instabilities respectively. The coupling constants of the quartic terms are positive, i. e. $u,\, v>0$, to ensure the stability of the system. Above action contains a dielectric part which is identical to one used in our study of quantum criticality in ferroelectrics in the reference \cite{I}.  The magnetic part of the action  is derived with the consideration of small ferro-magnetic component which can be integrated out, as well as the bipartite structure of the EuTiO$_3$ lattice that supports a Neel order below the transition point.  The detailed derivation of the anti-ferro magnetic part is given in references \cite{Sachdev, Aurbach}. In three dimension topological terms associated with quantum anti-ferromagnetic fluctuations are not important and hence neglected here. The coupling between the staggered-magnetization and the uniform magnetic field has some important consequences and hence its origin deserves some comments. At a microscopic 
level, an uniform magnetic field couples only to the uniform component of a Heisenberg spin. If we invoke a continuum description and using Haldanes' mapping\cite{Halden} integrate out the uniform component with a constraint of vanishing scalar product between the uniform and the staggered components, such term results.   Since in a quantum picture, statics and dynamics are coupled, an applied magnetic field induced precession of the magnetic vectors also play an important role in the study of phase transitions in quantum magnets. Now we start  with the following  mean field approximations.
\begin{eqnarray}
\vm(\vq, \omega)&=& m_0 \hat{z}\delta(\vq)\delta(\omega) + \vmp(\vq, \omega),\nonumber\\
<\vmp>&=&0,\, \, <\vp>=0
\end{eqnarray}
Here $\vm(\vq, \omega)$ is the Fourier transform of $\vm(\vx, \tau)$. The above approximations, along with a quasi-harmonic decoupling of the quartic terms, lead to the following mean field action,
\begin{eqnarray}
\mathcal{A}_{MF}&=&\frac{r_m}{2} m_0^2+\frac{v}{2}m^4_0\nonumber\\
&+&\int d^dq \frac{1}{\beta}\sum_n[(\frac{r_e}{2}+\omega_n^2+\frac{c_e q^2}{2}+  \frac{u}{2}\lambda_e)(\vp\cdot\vp) \nonumber\\&+&(\omega_n\vmp -i\vec{h}\times\vmp)^2 \nonumber\\
&+& (\frac{c_m q^2}{2} +\frac{r_m}{2} +\frac{v}{2}(2m_0^2+\lambda_m ))(\vmp\cdot\vmp)\nonumber\\
&-& \frac{w}{2}(m_0^2+\vmp\cdot\vmp)\vp\cdot\vp]
\end{eqnarray}
In the above expression, $\vec{A}\cdot\vec{A}= \vec{A}(\vq,\omega_n)\cdot\vec{A}(-\vq,\omega_n)$, $
\lambda_{(e,\,m)}=\int d^dq \frac{1}{\beta}\sum_n  \chi_{(e,\,m)}(\vq, \omega_n)$
and
\begin{eqnarray}
\chi_e(\vq, \omega_n)&=&<\vp(\vq,\omega_n)\cdot\vp(-\vq,\omega_n)>,\nonumber\\
 \chi_m(\vq, \omega_n)&=&<\vmp(\vq,\omega_n)\cdot\vmp(-\vq,\omega_n)> 
\label{S_mf}
\end{eqnarray}
{\bf Zero magnetic field ($h=0$):} In a zero external magnetic field, the 
self consistent equations for polarization and magnetic fluctuations are,
\begin{eqnarray}
\chi_e(\vq, \omega_n)&=&<\vp(\vq,\omega_n)\cdot\vp(-\vq,\omega_n)>\nonumber\\
&=&\frac{1}{\frac{r_e}{2}+\frac{c_e q^2}{2}+ \omega_n^2-\frac{w}{2}(m_0^2+\lambda_m)+\frac{u}{2}\lambda_e} 
\label{self_e},
\end{eqnarray}
and 
\begin{eqnarray}
\chi_m(\vq, \omega_n)&=&<\vmp(\vq,\omega_n)\cdot\vmp(-\vq,\omega_n)>\nonumber\\
&=&\frac{1}{\frac{r_m}{2}+\frac{c_m q^2}{2}+ \omega_n^2-\frac{w}{2}\lambda_e+\frac{v}{2}(2m_0^2+\lambda_m)}
\label{self_m}
\end{eqnarray}
respectively. The above two equations should be supplemented by the following expression for the {\it magnetic free energy} (within one loop correction) to determine $m_0$ in the magnetically ordered phase.
\begin{eqnarray}
 f_m &=&\frac{r_m}{2} (\vm_0)^2+\frac{v}{2}|\vm_0|^4-\frac{1}{2}Tr\log (\chi_m(\vq, \omega_n)) 
\label{fm}
\end{eqnarray}
We need to know  $m_0,\, \lambda_e,\,{\rm and}\, \lambda_m$ as a function of the temperature at various values of the system parameters, using equations [\ref{self_e}-\ref{fm}]. The extremization of $f_m$ with respect to $m_0$ gives,
\begin{equation}
r_m m_0 + 2v m_0^3 + vm_0\int d^dq \frac{1}{\beta}\sum_n \chi_m(\vq, \omega_n) =0
\end{equation}
Non-zero solution of $m_0$ reads as,
\begin{equation}
m_0^2 = \frac{-r_m -v\lambda_m}{2v} 
\end{equation}
This completes the description of a self-consistent formalism suitable for this system. Here we emphasize that at the critical value of the magnetic field, where the magneto-electric coupling is believed to be very small, the static dielectric constant for EuTiO$_3$ reaches a value $\mc{O}(10^2)$. Thus according to the classifications of various quantum paraelectrics the ferroelectric subsystem falls into the category of the {\it gapped quantum paraelectrics}, and is much more away from the ferroelectric quantum critical point than SrTiO$_3$. Such a dielectric state is characterized by a {\it crossover temperature} $T^*\sim \sqrt{r_e}$. Thus for pure EuTiO$_3$, we assume the ferroelectric crossover temperature $T^*$ is much higher than $T_N$, the Neel temperature. Low $T_N$ implies that, the system is in the vicinity of anti-ferromagnetic quantum critical point.  Thus the temperature dependence in the static dielectric constant at low temperature comes only from the magnetic fluctuations through magneto-
electric coupling. We consider the temperature dependence of $\lambda_m$ near as well as away from the magnetic critical point. Near a anti-ferro magnetic quantum critical point momentum cut-off becomes temperature dependent. Since in this material the dispersion relation for anti-ferromagnetic fluctuations is similar to that of the ferroelectric one the momentum cut-off at the critical point also $\sim T$. Thus within a non-self consistent estimate for the leading order temperature dependence we get,
\bea
\lambda_m\sim 
  \begin{cases}
    T\int_0^{T/\sqrt{c_m}} \frac{q^2dq}{c_mq^2} =  c_m^{-\frac{3}{2}}T^2  \, \text{near\, AFM-QCP }  \\
   T\int_0^\Lambda \frac{q^2dq}{r_m}=\frac{T\Lambda^3}{r_m} \sim T  \, \text{away from QCP } .
  \end{cases}
\eea
Above expression obtained using temperature dependent cut-off proposed by Mishra and Ramakrishnan\cite{SGM, sreeram} is similar to the thermal correction to the gap obtained later by Millis\cite{Hertz-millis}. Due to the magneto-electric coupling in this material, the above temperature dependence of $\lambda_m$ enters into the static dielectric constant and  results in the following temperature dependence of inverse static dielectric constant
\bea
 \chi^{-1}_e(0,0)\sim 
\begin{cases}
 \tilde{r_e} - wc_m^{-\frac{3}{2}}T^2\,\,({\rm near\, AFM-QCP})\\
\tilde{r_e} - \frac{w\Lambda^3}{r_m}T \,\,({\rm large} \, T_N).
\end{cases}
\eea
Here $\tilde{r_e}$ is the re-normalized value of $r_e$. Thus we see that, dielectric measurements can be considered as an indirect thermodynamic probe for magnetic systems in a magnetic quantum paraelectrics. It is to be noted that unlike the similar discussions in our second chapter, we do not equate $c_m$ to unity here. We will see in the subsequent discussions that along with quartic couplings and magneto-electric couplings, $c_m/c_e$ will be an important parameter to determine the dominance between the magnetic fluctuations and the paraelectric fluctuations to contribute to  the static dielectric susceptibility of this material in certain appropriate circumstances. In next subsection we will consider change in the dielectric behavior in this system in case of field induced transition in the magnetic subsystem.

 {\bf Non-zero magnetic field ($h\neq 0$):} Non-zero $h$ modifies anti-ferromagnetic order, develops a ferromagnetic order along the direction of the field and the resulting magnetic configuration becomes {\it canted}\cite{Sachdev}. Firstly it is apparent from our starting action (eqn.(\ref{S})) that in case of a 
non zero $h$ along the z-direction, anti-ferromagnetic gap in the transverse plane (with respect to the field) changes to,
\be
  r_m\sim r_m(0)-h^2
\ee
Where $r_m(0)<0$ is the value of $r_m$ at zero magnetic field. Thus $h$ reduces the gap in the transverse directions. In the regime $r_m<0$ and $|r_m|>h^2$, $m_0$ is still non-zero and it increases with increasing $h$ in the following manner,
\be
m_0\sim (h^2-r_m(0))^{\frac{1}{2}} \sim (h-h_0)^{\frac{1}{2}}
\label{mafm}
\ee
Where $h_0=\sqrt{r_m(0)}$. But the ferromagnetic order along the direction along the field grows more rapidly with applied magnetic field as follows,
\be
m_{fm} =-\frac{\partial F}{\partial h}\sim \frac{h(h^2-r_m(0))}{v} 
\label{mfm}
\ee 
A schematic phase diagram for field induced transition in the magnetic subsystem is shown in figure (\ref{cant}). In our case $r_m(0)$ is negative and we consider the external magnetic field induced modification of the anti-ferromagnetic -ground state to a canted state with partial ferromagnetic order and its effect on the static dielectric susceptibility.  An experimentally observed fact is that the increase in anti-ferromagnetic  component results in the suppression of dielectric constant while the roll of the uniform component is just opposite to it. If we assume both the components couple to the polarization in the same fashion, we can make an estimate of the critical value of the magnetic field ($h_c$) which exactly nullifies the effects of magnetic order on the static dielectric constant, in the following way. Using eqn. (\ref{mafm}) and (\ref{mfm}) we get,
\bea
 (h_c^2-r_m(0))&=&\frac{1}{c}\times h^2_c(h_c^2-r_m(0))^2\nonumber\\
\Rightarrow h_c&=&\frac{r_m(0)\pm\sqrt{r_m^2(0)+4c}}{2}
\eea
Where $c$ is a non-universal constant and so is $h_c$. Thus at $r_m(0)=0$ i. e. at anti-ferromagnetic quantum critical point, $h_c\sim \sqrt{c}$. Apart from this, external magnetic field has one more effect on quantum criticality. 
\figcant
In case of field induced transition, the finite temperature behavior near quantum critical point will also be different. If we look back the action(\ref{S}), we see that the magnetic field adds a new dynamic term $\sim -i\vec{h}\times \vmp\cdot \partial_{\tau}\vmp$ which is linear in $\omega_n$. Thus the dynamic exponent $z=2$  and the temperature dependent momentum cut-off for magnetic excitations $\Lambda\sim \sqrt{\frac{T}{hc_m}}$ in this case. For small $h$, i. e. when $m_{fm}<< m_0$, 
\be
\lambda_m\sim  T\int_0^{\sqrt{T/hc_m}} \frac{q^2dq}{c_mq^2} =  c_m^{-\frac{3}{2}}h^{-\frac{1}{2}}T^{3/2}.
\ee
Thus one would expect a $T^{3/2}$ contribution, from the magnetic subsystem to the inverse static dielectric susceptibility at low temperature near anti-ferromagnetic quantum critical point. Here we assume that the applied field is  small enough to induce a meta-electric transition. 
\figchic
\figchi

{\bf Near ferroelectric quantum critical point:} So far, we have considered the dielectric subsystem as a spectator with a temperature independent dielectric susceptibility at low temperature. However, one can
make $T^*$ closer or smaller than $T_N$ through doping. A generic possibility is replacing O$^{16}$ by O$^{18}$ in EuTiO$_3$, as is done in case of SrTiO$_3$\cite{O18}. Such a doping can create a reduced crossover temperature $\tilde{T}^*(x)\sim (1-x)^{\frac{1}{2}} T^*$, (where $x$ is the impurity concentration) and move the system towards ferroelectric quantum critical point without affecting the magnetic subsystem. At finite temperature near ferroelectric quantum critical point, dipolar contribution to the inverse static dielectric constant is $\sim uc_e^{-\frac{3}{2}} T^2$ which will compete with negative contribution ($\sim -T^{\nu}, \, \nu=(1,\, 2,\, 3/2)$), coming from the coupling with magnetic subsystem. If we assume that these two quantum critical point do not affect each other, then considering the leading order temperature dependence to the static dielectric susceptibility, we can write,
\be
\chi_e^{-1} = \alpha+\gamma_e T^2-\gamma_\nu T^{\nu} 
\ee
Where $\alpha,\, \gamma_e,\, \gamma_\nu\, {\rm with} \, \gamma_e,\, \gamma_\nu>0 $ are constant which varies from system to system and $\gamma_e$ and $\gamma_\nu$ are proportional to $u c_e^{-\frac{3}{2}}$ and $w$ respectively. For different values of $\nu$, $\gamma_\nu$  is given as follows.
\bea
\gamma_\nu\sim 
\begin{cases}
 w\frac{\Lambda^3}{r_m}\,\,\text{for $\nu=1$}\\
w c_m^{-\frac{3}{2}}h^{-\frac{1}{2}}\,\,\text{for $\nu=3/2$}\\
wc_m^{-\frac{3}{2}}\,\,\text{for $\nu=2$}\\
\end{cases}
\eea
 Among all these values, except $\gamma_{3/2}$ other $\gamma_\nu$s are fixed by system parameters and can not be controlled externally. Since $\gamma_{3/2}$ depends on the external magnetic field, the temperature scale up-to which a  $T^{3/2}$ behavior of the dielectric susceptibility should be observed is also depends on it and can be tuned externally in an experimental situation. However for $\nu =1 \,{\rm or}\, 3/2$, the temperature dependence of static dielectric susceptibility at low temperature will be dominated by anti-ferromagnetic quantum critical point with  ($1/T^{\nu}$ increase) and there will be a maxima at a temperature $T_{max}= (\frac{2\gamma_e}{\nu\gamma_\nu})^{\frac{1}{\nu-2}}$ as shown in figure \ref{chic}.
For $\nu =2$, there will be a competition between anti-ferromagnetic quantum critical point and ferroelectric quantum critical point and depending on the values of $\gamma_2/\gamma_e$ an anti-ferromagnetic quantum critical point dominated behavior with $1/T^{2}$ increase or a ferroelectric quantum critical point dominated behavior with $1/T^{2}$ decrease can be found in the static dielectric constant as shown in the figure \ref{chi}.
\section{Discussion}
In this work we present a mean field theory to discuss the temperature and the magnetic field dependence of the static dielectric susceptibility of a magnetic quantum paraelectric at low temperature.
In this material anti-ferromagnetic fluctuations are coupled to the polarization fluctuations and their interplay can lead to many interesting thermodynamic consequences when some non-thermal control parameters of both  fluctuations are tuned to near critical values. We focus on the behavior of the system in the vicinity of two such quantum critical points both in absence and in presence of an external magnetic field. Based on scaling argument near quantum critical points, we predict  that there is a possibility that the low temperature suppression of the  static dielectric susceptibility  due to magnetic order can be compensated by polarization fluctuations and the  static dielectric susceptibility would take a $1/T^2$ form as predicted for quantum critical ferroelectrics\cite{I}.  On the other hand because of magneto-electric coupling there is a possibility of new power law behavior of the  static dielectric susceptibility in presence of an external magnetic field and is predicted to be $1/T^{3/2}$ in this 
case.  At present, up-to our knowledge, there is no report on experimental investigations on the simultaneous effects of two such quantum critical points. Hence fitting some experimental data through the  numerical solutions of self-consistent equations is not tried here. Rather possible new features in this multi-ferroic material near various quantum critical points are explored. 
 This system, in many aspects is similar to the systems where anti-ferromagnetic  order parameter is coupled to superconductivity\cite{Arovas} and can be considered as a case for multiferroic quantum criticality\cite{spalek}. 
 \section*{Acknowledgment}The author would like to thank  S. G. Mishra for many useful discussions. 

\end{document}